\documentclass[aps,twocolumn,preprintnumbers,amssymb,amsfonts,latexsym,groupedaddress,superscriptaddress]{revtex4}

\usepackage{graphicx}
\usepackage{subfigure}
\usepackage{bm}% bold math
\usepackage{amsmath}
\usepackage{color}
\usepackage{diagbox}

\usepackage{indentfirst}

\newcommand{\bmath}{\begin{mathletters}}
\newcommand{\emath}{\end{mathletters}}
\newcommand{\be}{\begin{eqnarray}}
\newcommand{\ee}{\end{eqnarray}}
\newcommand{\ba}{\begin{array}}
\newcommand{\ea}{\end{array}}

\newcommand{\calK} {\mathcal K}

\newcommand{\rmR} {{\mathrm{R}}}
\newcommand{\rmX} {{\mathrm{X}}}

\newcommand{\rmS} {\mathrm{S}}
\newcommand{\rmI} {\mathrm{I}}
\newcommand{\rmB} {\mathrm{B}}

\begin{document}

\title{Unusual Transport Properties with Non-Commutative System-Bath Coupling Operators}

\author{Chenru Duan}
\affiliation{Department of Chemistry, Massachusetts Institute of Technology,
77 Massachusetts Avenue, Cambridge, MA 02139}
\affiliation{Singapore-MIT Alliance for Research and Technology (SMART) Centre, Singapore 138602}
\author{Chang-Yu Hsieh}
\affiliation{Singapore-MIT Alliance for Research and Technology (SMART) Centre, Singapore 138602}
\author{Junjie Liu}
\affiliation{Singapore-MIT Alliance for Research and Technology (SMART) Centre, Singapore 138602}
\author{Jianshu Cao}
\email{jianshu@mit.edu}
\affiliation{Department of Chemistry, Massachusetts Institute of Technology,
77 Massachusetts Avenue, Cambridge, MA 02139}
\affiliation{Singapore-MIT Alliance for Research and Technology (SMART) Centre, Singapore 138602}
\affiliation{Beijing Computational Science Reseach Center, Beijing 100084, China}

\begin{abstract}
Understanding non-equilibrium  heat transport is crucial for controling heat flow in nano-scale systems.
We study thermal energy transfer in a generalized non-equilibrium spin-boson
model (NESB) with non-commutative system-bath coupling operators and discover unusual transport properties.
Compared to the conventional NESB, the heat current is greatly enhanced by rotating the coupling operators. 
Constructive contribution to thermal rectification can be optimized when two sources of asymmetry,
system-bath coupling strength and coupling operators, coexist.
At the weak coupling and the adiabatic limit, the scaling dependence of heat current on the coupling
strength and the system energy gap changes drastically when the coupling operators become non-commutative.
These scaling relations can further be explained analytically by the non-equilibrium polaron-transformed Redfield equation.
These novel transport properties, arising from the pure quantum effect of non-commutative coupling operators, 
should generally appear in other non-equilibrium set-ups and driven-systems.

\end{abstract}

\maketitle

% \section{Introduction}
\textit{Introduction.} With the advance of nano-scale quantum technologies and global efforts on sustainable development, understanding
the fundalmental laws of heat transport at the microscopic level has attracted much theoretical and expiremental
attention~\cite{Ono2002:Science,Gotsmann2017:NatNano,Plenio2013:PRL,Ventra2011:RMP,Singer2016:Science,Leitner:2015}. 
Treated as a minimal model for anharmonic molecular junctions, the non-equilibrium spin-boson 
model (NESB) has been extensively investigated with various theoretical~\cite{
Segal2011:PRB,Segal2014:JPCA,Segal2017:NJP,Ojanen2011:PRB,JSWang2014:JCP,Junjie2017:PRE,Esposit02015:PRL,
Cao2015:SciRep,Cao2016:NJP,Cao2017:PRA,Changyu2017}
and numerical methods~\cite{Saito2013:PRL,Tanimura2015:JCP,Tanimura2016:JCP,HBWang2008:CPL,Cao2016:FP,Javier2017:PRB,Qshi2017:PRB}. 
Many interesting propeties of heat transfer have been found, including a turnover of the heat current
on a function of system-bath coupling strength~\cite{HBWang2008:CPL,Cao2015:SciRep}.

Symmetry lies in the center in physical science and plays a crucial role in heat transport problems
~\cite{Gross1996:PNAS,Hanggi2014:PhyRep,Walschaers2013:PRL,Cao2016:SciRep,Nitzan2002:PRL}.
In the NESB, the asymmetry of the system-bath coupling strength leads to thermal rectifcation
~\cite{Segal2005:PRL,BWLi:2012:RMP}.
For a two-bath spin-boson model at zero temperature, an additional competition arises due to the non-commutative (asymmetric)
system-bath coupling operators, and the effect of ``frustration of decoherence'' emerges
~\cite{Kohler2013:PRB,Vojta2012:PRL,Vojta2014:PRB,Afﬂeck2005:PRB,Afﬂeck2003:PRL}.
Despite the huge effort on studying their equilibrium state, little attention has been paid to investigate
heat transport under the influence of non-commutative coupling operators.
In a recent paper~\cite{Tanimura2016:JCP}, Kato and Tanimura found the
``correlation among system-bath interaction'' effect on the heat current.
Since they mainly focused on the definition of heat current from the bath prespective,
the unusual transport phenomena arising from non-commutative coupling operators have not been explored.

In this paper, we thoroughly investigate the influence of non-commutative coupling operators on heat transport
in the NESB using a numerically accurate method, extended hierarchy equation of motion (HEOM)
~\cite{Tanimura1989:JPSJ,Tanimura2006:JPSJ,ZFTang2015:JCP,Duan2017:PRB}. 
In the weak coupling regime,
a smooth transition in the scaling of the steady state heat current from $I \sim \alpha$ to $I \sim \alpha^{2}$ is found 
as we rotate the system-bath coupling operators.
For a generalized NESB with non-commutative coupling operators (nc-NESB), 
a plateau for heat current occurs as $\Delta/\omega_{c} \rightarrow 0 $,
while the heat current drops to zero when $\Delta/\omega_{c} \rightarrow 0 $ in the conventional NESB (c-NESB). 
These observations can be quantitatively explained by the non-equilibrium 
polaron-transformed Redfield equation (NE-PTRE)~\cite{Cao2015:SciRep,Cao2016:NJP,Cao2017:PRA,Changyu2017}.
The steady-state heat current of nc-NESB is significantly enhanced compared 
to the c-NESB by simply rotating the coupling operators between the system and bath.
In addition, two sources of asymmetry, namely, asymmetric coupling strength and non-commutative coupling operators,
can contribute constructively to thermal rectification, giving rise to a larger rectification ratio than
cases with only one source of asymmetry.

% \section{Model}
\textit{Model and methods.}
The Hamiltonian for a generalized NESB is
\begin{eqnarray}
&&H = H_\rmS + H_{\rmB,1} +  H_{\rmB,2} +V_1 \otimes B_1 + V_2 \otimes B_2 \nonumber\\
&&=\Delta \sigma_z + \sum\limits_{\nu=\{1,2\},j} \omega_{\nu,j} b_{\nu,j}^\dagger b_{\nu,j} +
\sigma_x \sum\limits_j g_{1,j} (b_{1,j}^\dagger  \nonumber\\
&&+b_{1,j}) + \sigma_\theta \sum\limits_j g_{2,j} (b_{2,j}^\dagger+ b_{2,j}).
\label{eq_n01}
\end{eqnarray}
Here $\sigma_{i}$ ($i = x, y, z$) denotes the Pauli matrices, $\Delta$ is the half enengy gap of the two-level system,
and $b_{\nu, j}^{\dagger}$ ($b_{\nu,j}$) is the creation (annihilation) operator of the $j$-th harmonic oscillator
in the $\nu$-th bosonic bath. 
We consider the effect of non-commutative coupling by introducing a
parameter $\theta$ for the coupling operator between the system and the second bath,
i.e. $\sigma_\theta = \sigma_{z} \cos\theta + \sigma_{x} \sin\theta$, so that it can point at any direction
on the $x-z$ plane of a Bloch sphere. 
Due to the rotational symmetry of the model, we can restrict our study to
$0 \le \theta \le \pi/2$ without loss of generality. 
Note that our Hamiltonian reduces to the c-NESB at $\theta = \pi/2$, othewise it represents a nc-NESB.

The dissipative effect on the system can be characterized by a spectral density 
$J_{\nu}(\omega) = 4 \pi \sum_j g_{\nu,j}^2 \delta (\omega - \omega_{\nu,j}) = \pi \alpha_{\nu} \omega^s \omega_c^{1-s}
f(\omega/\omega_c)$, which is defined by the dimensionless system-bath coupling strength $\alpha_{\nu}$, the cutoff 
function of the environment $f(\omega/\omega_c)$, and the spectral exponent $s$ that catagorizes bath into
sub-Ohmic ($s < 1$), Ohmic ($s = 1$) and super-Ohmic ($s > 1$). Throughout this paper, we choose a super-Ohmic
sepctral exponent $s = 3$ and a rational cutoff function $f(\omega/\omega_c) = 1/(1+ (\omega/\omega_c)^2)^{4}$ for
both high and low temperature baths and assume $\alpha_{1} = \alpha_{2} = \alpha$ unless specified.
The atomic unit $\hbar = k_B = 1$ is used and the bath cutoff frequency is treated as an energy unit ($\omega_{c}=1$).
Further, the bath correlation function, $C_{\nu}(t) = 1/\pi \int_{0}^{\infty}J_{\nu}(\omega)
\left[\coth\frac{\beta_{\nu}\omega}{2}\cos\omega t-i\sin \omega t\right]d\omega$, 
with the inverse temperature $\beta_{\nu}={1} / {T_{\nu}}$,
uniquely determines the bath properties and their influence on the system.

% \section{extended hierarchy equation of motion}

Due to its numerical accuracy in propagating the dynamics and its camptibility for heat current calculation, 
the HEOM has become a popular numerical methods for simulating 
heat transport problems~\cite{Tanimura1989:JPSJ,Tanimura2015:JCP,Tanimura2016:JCP,Qshi2017:PRB,Javier2017:PRB}. 
In this paper we adopt the extended HEOM which can be applied to more general bosonic baths than the Debye-Lorentz form
~\cite{ZFTang2015:JCP, Duan2017:PRB, Duan2017:JCP, Cao2018:JCP}.
In the extended HEOM,
bath correlation functions and their time derivatives are decompsed by some finite basis sets $\{\phi_{\nu,j}^ \rmX(t)\}$ 
where $C_{\nu}^\rmX(t) = \sum_j a_{\nu,j}^\rmX \phi_{\nu,j}^\rmX(t)$ and
$\frac \partial {\partial t} C_{\nu}^\rmX(t)= \sum_{j,j^\prime} a_{\nu,j}^\rmX \eta_{\nu,j,j^\prime}^\rmX \phi_{\nu,j^\prime}^\rmX(t)$.
Here, $X= R$ or $I$ denotes the real (imaginary) part of the bath correlation function, $C_\nu(t)$.
Based on those closed funtion sets $\{\phi_{\nu,j}^ \rmX(t)\}$,
auxiliary fields $\vec{\sigma}(t)$ can be constructed and their evolutions are expressed in a time-local form~\cite{Tanimura2006:JPSJ,SM},
\begin{eqnarray}
\frac {\partial} {\partial t} \vec{\sigma}(t) = \vec{\calK}\ \vec{\sigma}(t).
\label{eq_n02}
\end{eqnarray}
Combined with the full counting statistics~\cite{Esposito2009:RMD}, we can express the heat current from a bath perspective
with our first-order auxiliary fields~\cite{Tanimura2016:JCP,Javier2017:PRB,SM} as
\begin{eqnarray}
&&I_{\nu}(t) = - \sum\limits_{j,j^\prime} a_{\nu,j}^\rmR \eta_{\nu,j,j^\prime}^\rmR 
\sigma_1^{\vec{n}_\nu = (j^\prime)}(t) \nonumber\\
&&- \sum\limits_{j,j^\prime} a_{\nu,j}^\rmI \eta_{\nu,j,j^\prime}^\rmI  \sigma_1^{\vec{m}_\nu = (j^\prime)}(t),
\label{eq_n03}
\end{eqnarray}
where the steady state heat current is obtained as $t \rightarrow +\infty$. 
Unlike some methods which are restricted by the system-bath coupling operators, 
Eq.~(\ref{eq_n03}) can be directly applied to calculate the steady-state heat current ($I$) for a nc-NESB.

% \section{Scaling relation of Heat Current}
\begin{figure}[t!]
	\includegraphics[width=1.00\columnwidth]{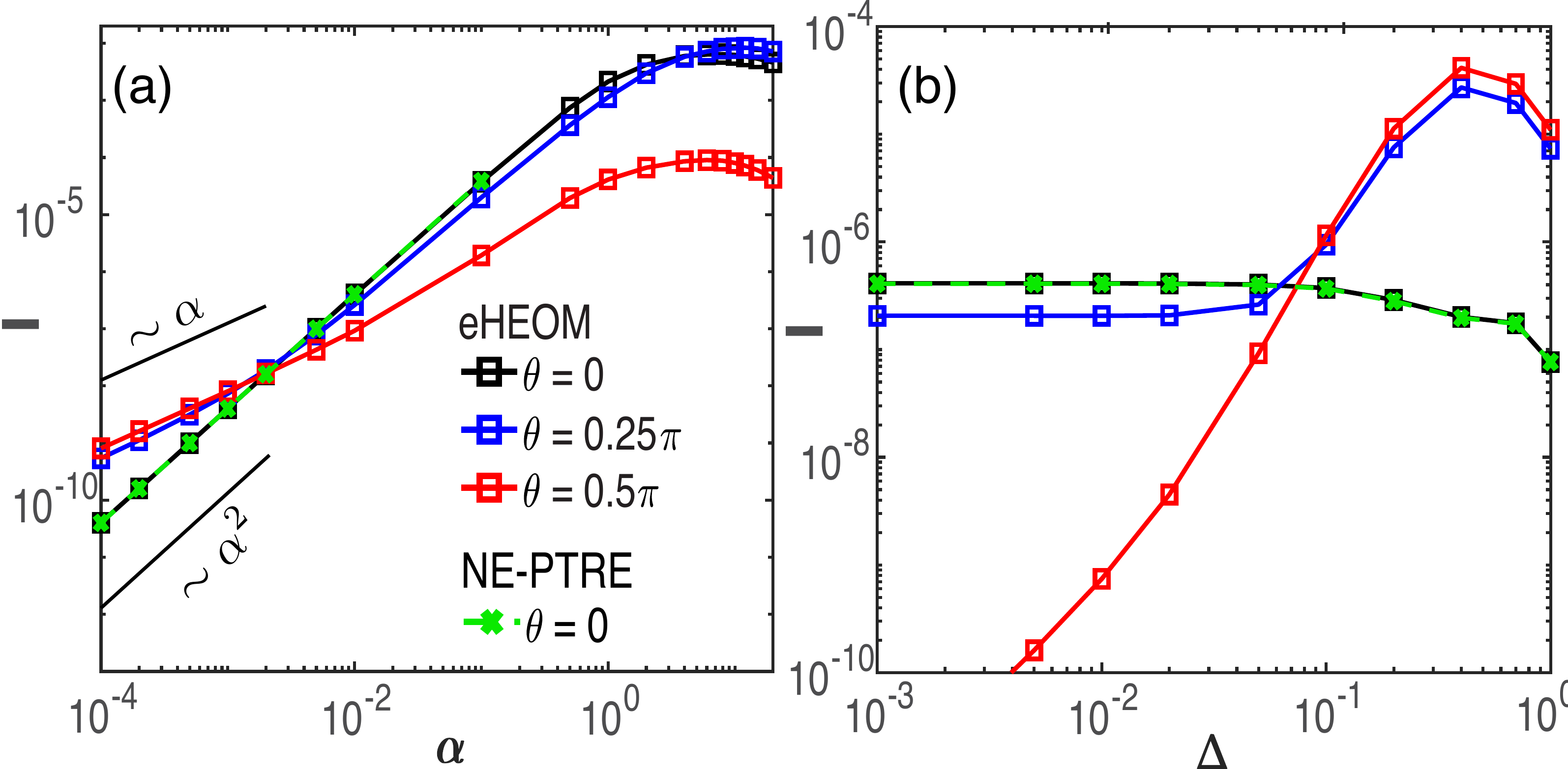}
	\caption{Heat current $I$ as a function of (a) coupling strength 
	 $\alpha$  at $\Delta = 0.05$ and (b) half energy gap $\Delta$ at $\alpha=0.01$ with different coupling operators for
	the second bath: $\theta = 0$ (black and green), $\theta = 0.25\pi$ (blue) and $\theta=0.5\pi$ (red). Squeares are
	results from the extended HEOM and crosses are results obtained by the NE-PTRE. Other parameters are $T_{1} = 1$ and
	$T_{2} = 0.9$.}
	\label{fig_n01}
\end{figure}

\textit{Scaling relation of heat current.}
Figure~\ref{fig_n01} (a) demonstrates the relationship between the steady state heat current 
and the coupling strength for different nc-NESB configurations ($\theta$). 
In the weak coupling regime,  the heat current of c-NESB ($\theta=0.5\pi$) is
proportional to the coupling strength, $I \sim \alpha$, which agrees with the Redfield equation. 
While for a nc-NESB, this scaling behavior is altered. 
At the extreme case when the two coupling operator are orthogonal ($\theta = 0$), we observe $I \sim \alpha^{2}$. 
For $0 < \theta < 0.5\pi $, there is a smooth transition from $I \sim \alpha^{2}$ to $I \sim \alpha$~\cite{SM}. 
This continuous transition is implied by the heat current expression in the Heisenberg picture~\cite{Tanimura2016:JCP},
\begin{eqnarray}
&&I = \left\langle \left[\Delta \sigma_z, (\sigma_z \cos\theta  + 
\sigma_x \sin\theta)\otimes \hat{B}_2\right] \right\rangle \nonumber\\
&&+\left\langle \left[\sigma_x\otimes \hat{B}_1, (\sigma_z \cos\theta  + 
\sigma_x \sin\theta)\otimes \hat{B}_2\right] \right\rangle  
\label{eq_n04}
\end{eqnarray}
where $\langle \ldots \rangle$ is the trace of the steady state total density matrice over all degrees of freedom
and $\hat{B}_{\nu}$ denotes the bath operator in the Heisenberg picture.
In the weak coupling limit, the first term in
Eq.~(\ref{eq_n04}) gives linear dependece of $I$ on $\alpha$ but vanishes at $\theta=0$, where the second term
predicts $I \sim \alpha_1 \alpha_2 \sim \alpha^{2}$.
As shown in Fig.~\ref{fig_n01} (a), despite the difference in the scaling relation at small $\alpha$ when varying $\theta$,
the heat currents all show a turnover behavior,
which indicates the system-bath inseparatability for strong system-bath interaction~\cite{Alan2015:JPCL}. 
Nevertheless, significant enhancement in the heat current can still be observed for nc-NESB 
($\theta\neq 0.5\pi$) compared to that of c-NESB ($\theta = 0.5\pi$) except for a very weak interaction strength.

For different NESB configurations, the $\Delta$ dependence on the heat current $I$ is depicted in
Fig.~\ref{fig_n01} (b). For the c-NESB, $I$ drops to zero as $\Delta \rightarrow 0$. For a nc-NESB, a 
plateau for the heat current appears when $\Delta$ approaches zero. This phenomena can
also be explained by Eq.~(\ref{eq_n04}) in which the first term depends explicitly on $\Delta$ while the second term does not,
thus a non-zero heat current can still arise at $\Delta = 0$ for a nc-NESB.
Physically, at $\Delta = 0$ the total Hamiltonian
can simply be diagonalized for $\theta = 0.5\pi$ by a full polaron transfomation, where the orthogonality catastrophe
prevents any channels for heat transfer\cite{Anderson1967:PRL}.
However, this orthogonality catastrophe does not occur when $\theta \neq 0.5 \pi$, and those non-diagonal parts give rise
to a non-zero heat current. Note that the Redfield equation cannot capture 
the second term in Eq.~(\ref{eq_n04}), which is due to high order system-bath interaction.
Therefore, it requires other methods to evaluate higher order interaction. 
So we introduce the non-equilibrium polaron-transformed Redfield equation (NE-PTRE) below
~\cite{Cao2015:SciRep,Cao2016:NJP,Cao2017:PRA}.

To develop a clear physical picture, we consider a specific configuration, $\theta = 0$, 
i. e., the two system-bath coupling operators are orthogonal,
so that the first term in Eq.~(\ref{eq_n04}) vanishes. 
With a full polaron transformation of the second bath and the introduction of the counting field $\chi$
on the first bath, we obtain the transformed Hamiltonian $H^{\prime}$ as~\cite{SM},
\begin{eqnarray}
&&H^\prime = \Delta \sigma_z + \sum\limits_{\nu,j} \omega_{\nu,j} b_{\nu,j}^\dagger b_{\nu,j} 
+ (\sigma_x \cosh 2A_2 \nonumber\\
&&+ i \sigma_y \sinh 2A_2) \sum\limits_j g_{1,j}(b_{1,j}^\dagger[\frac \chi 2] + b_{1,j}[\frac \chi 2]),
\label{eq_n05}
\end{eqnarray}
where $A_{2} = \sum_{j}g_{2,j}/\omega_{2,j} (b_{2,j}^\dagger - b_{2,j})$ and $O[\chi] = 
\exp(i \chi \sum_j \omega_{1,j} b_{1,j}^\dagger b_{1,j}) O \exp(-i\chi \sum_j \omega_{1,j} b_{1,j}^\dagger b_{1,j})$.
Following the standard procedure of the NE-PTRE~\cite{Cao2015:SciRep,Cao2016:NJP,Cao2017:PRA} and a perturbation expansion
on $\alpha$, the heat current can be obtained as~\cite{SM},
\begin{eqnarray}
&&I = -2 \int\limits_0^\infty dt (C_{1}^\rmR(t) \dot Q_{2}^\rmI(t) + C_{1}^\rmI(t) \dot Q_{2}^\rmR(t) ) \cos 2 \Delta  t    \nonumber\\
&& +2\xi(\Delta) \int\limits_0^\infty dt (C_{1}^\rmI(t) \dot Q_{2}^\rmI(t) + C_{1}^\rmR(t) \dot Q_{2}^\rmR(t) ) \sin 2 \Delta t, 
\label{eq_n06}
\end{eqnarray}
where $\xi(\Delta) = {\int_0^\infty dt  C_{1}^\rmI \sin 2 \Delta t}/ {\int_0^\infty dt  C_{1}^\rmR \cos2 \Delta t}$ is independent
of $\alpha$ and $Q_2(t) = Q_{2}^\rmR(t)+ i Q_{2}^\rmI(t) =  2 \int_0^\infty d\omega J_2(\omega) 
(n_2(\omega)\exp(i \omega t) +(n_2 (\omega) +1)\exp(-i \omega t))/\omega^2$.
Here we have $\dot  Q_{2}^\rmX(t) = dQ_{2}^\rmX(t)/dt$ and the Bose-Einstein distribution function
$n_{\nu}(\omega) = 1/(\exp(\beta_{\nu}\omega) - 1)$.
On one hand, both $C_{1}^\rmX(t)$ and $Q_{2}^\rmX(t)$ are linearly dependent on the coupling strength $\alpha$, giving
$I \sim \alpha^{2}$. On the other hand, Eq.~(\ref{eq_n06}) clearly predicts a non-vanishing heat current
$I(\Delta=0) = -2 \int_0^\infty dt (C_{1}^\rmR(t) \dot Q_{2}^\rmI(t) + C_{1}^\rmI(t) \dot Q_{2}^\rmR(t)$.
In the adiabatic limit of $\Delta \ll 1$, we have $I(\Delta \ll 1) - I(\Delta=0)  \sim \Delta ^2$, which explains the plateau. 
As shown in Fig.~\ref{fig_n01}, results obatined by Eq.~(\ref{eq_n06}) are in excellent agreement with those of the extended HEOM.
Interestingly, our heat current expression (Eq.~(\ref{eq_n06})) is not limited to small $\Delta \ll 1$, as it does not involve
perturbative expansion of $\Delta$, which was an issue of the NE-PTRE but recently improved~\cite{Changyu2017}.
However, since only the second bath is displaced in our polaron transformation, the NE-PTRE cannot be applied to the
entire regime of coupling strength.

% \section{Heat Current Optimization}
\begin{figure}[t!]
	\includegraphics[width=1.00\columnwidth]{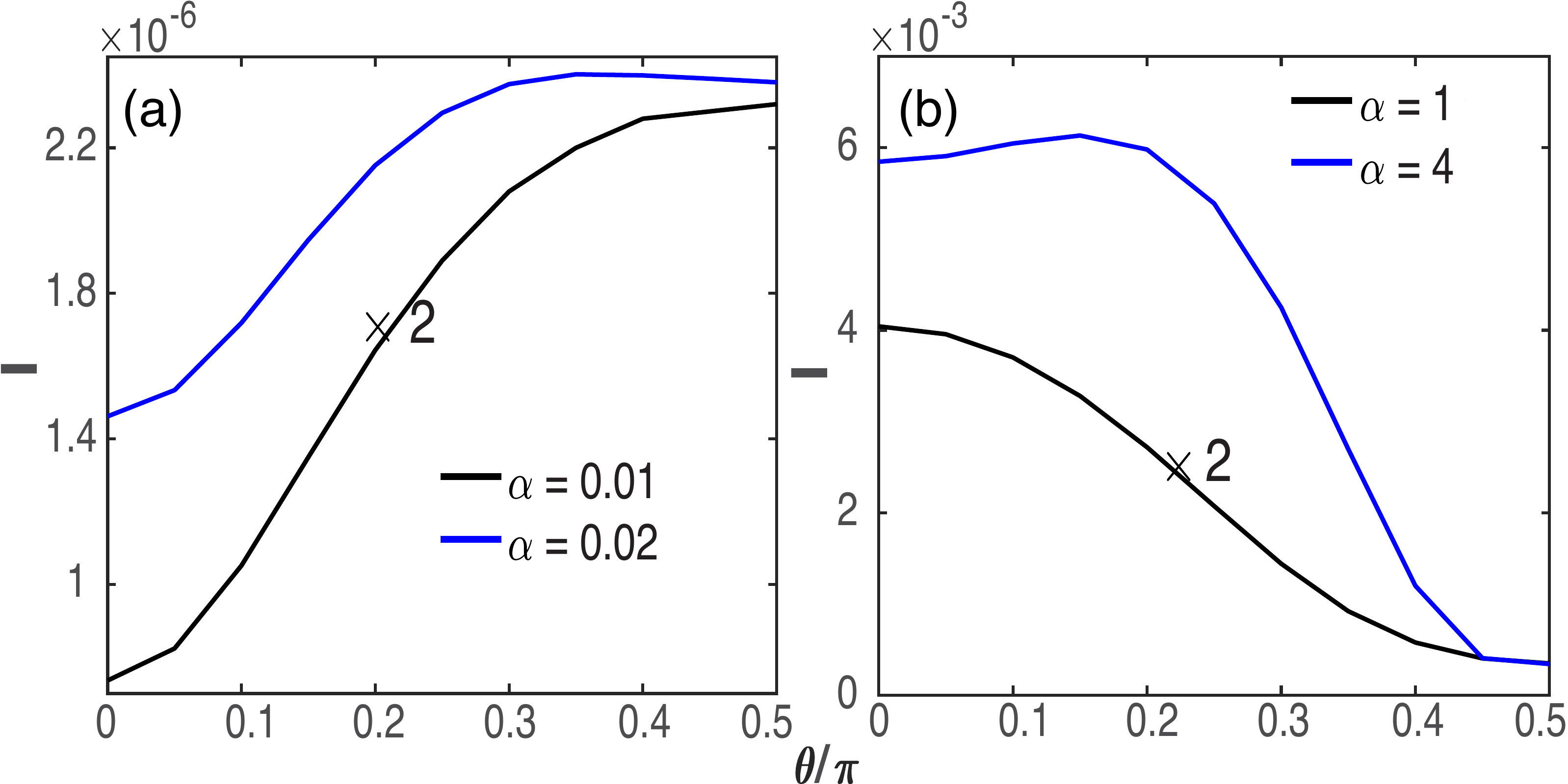}
	\caption{Heat current as a function of the second operator direction $\theta$.
	(a) $\alpha = 0.01$ (black), $\alpha = 0.02$ (blue) and (b) $\alpha = 1$ (black), $\alpha = 4$ (blue).
	Black lines in both (a) and (b) are multiplied by a factor of $2$ for a better view of results.
	Other parameters are $T_{1} = 1$, $T_{2} = 0.9$ and $\Delta=0.1$.}
	\label{fig_n03}
\end{figure}

\textit{Optimization of heat current.} 
Optimal thermal properties are always of great interest to the performance of molecular junctions,
quantum heat engines and heat pumps~\cite{Ventra2011:RMP}.
In our model, heat current can be optimized with respect to $\theta$, given that other parameters,
$\alpha,\Delta,T_{1}$ and $T_{2}$ are fixed.
Figure \ref{fig_n03} demonstrates the heat current as we rotate the second coupling operator 
from $\sigma_z$ to $\sigma_{x}$ direction at a fixed energy gap $\Delta = 0.1$. 
Various behaviors are obsersed. At a very weak coupling strength of $\alpha=0.01$, the
heat current grows monotonously as $\theta$ increases from $0$ to $0.5\pi$~\cite{Tanimura2016:JCP}. 
On the contrary, for $\alpha = 1$, the heat current decreases monotonically with increasing $\theta$.
Non-monotounous $\theta$ depedence emerges for $\alpha = 0.02$ and $\alpha = 4$,
where the heat current is maximal at an intermediate configuration, i.e. $0 <\theta_{\mathrm{opt}} < \pi/2$.

\begin{figure}[t!]
	\includegraphics[width=1.00\columnwidth]{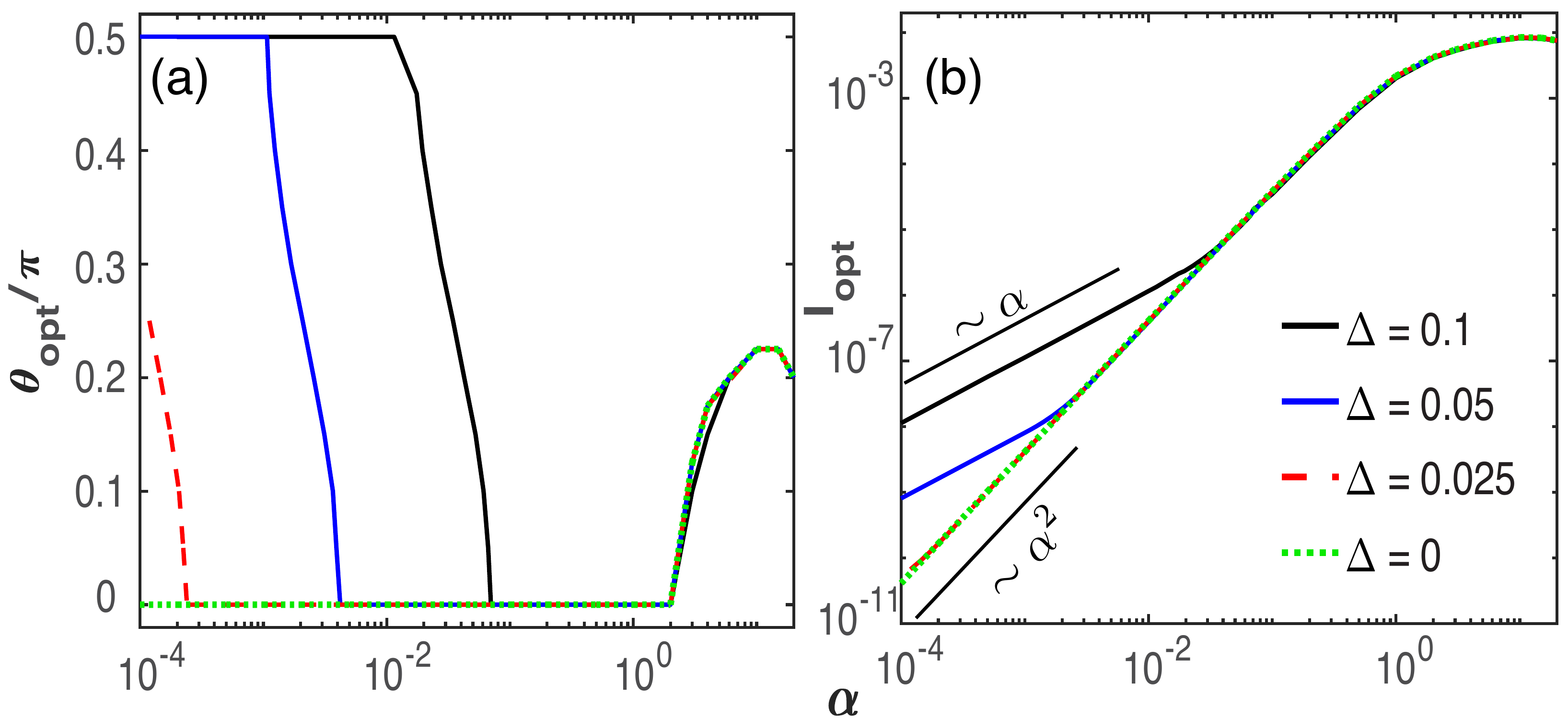}
	\caption{The coupling strength dependence of the optimized (a) angle and (b) heat current 
	for a series of $\Delta$ at the scaling limit:
	$\Delta=0.1$ (black soilid line), $\Delta=0.05$ (blue solid line),  $\Delta=0.025$ (red dashed line)
	and $\Delta=0$ (green dotted line). Temperatures for two baths are $T_{1} = 1$ and $T_{2} = 0.9$ respectively.}
	\label{fig_n02}
\end{figure}

To develop a better understanding, we further study the relationship between the interaction strength and
the optimal angle $\theta_\mathrm{opt}$ at which the heat current reaches its maximum value $I_\mathrm{opt}$.
Results are shown in Fig.~\ref{fig_n02} (a).
For a finite $\Delta$, four distict regimes can be identified over the range of
the coupling strength under investigation. 
(I) For a very weak system-bath coupling, the c-NESB ($\theta=0.5\pi$) gives the maximal heat current 
as the linear term in Eq.~(\ref{eq_n04}) is dominant in comparison with the second order term.
(II) A transition of the optimal angle from $\theta_\mathrm{opt}=0.5\pi$ to $\theta_\mathrm{opt}=0$
follows with the increasing interaction strength,
because of the non-negligible contribution from the second order term in Eq.~(\ref{eq_n04}).
(III) The effect of second order heat current is prominant within a certain range of $\alpha$ 
where the optimal angle stays at $\theta_\mathrm{opt}=0$. 
(IV) Contribution of even higher order transport processes gradually intervene and eventually become dominant at
very strong coupling strength, so that $0< \theta_\mathrm{opt} < 0.5\pi$ can be observed. 
These four regimes are also indicated in Fig.~\ref{fig_n02} (b),
which depicts the relationship between $\alpha$ and $I_\mathrm{opt}$: $I_\mathrm{opt}
\sim \alpha$ at the very weak interation (I) followed by a transition (II) to $I_\mathrm{opt} \sim \alpha^{2}$ (III)
at the intermediate coupling strength. As the system-bath interaction keeps increasing, $I_\mathrm{opt}$ deviates from
the $\alpha^{2}$ dependence and a turnover appears (IV).
Although this turnover behavior is inevitable due to the inseperatibilty between system and bath,
the maximum heat current, $I_{\mathrm{max}} = \mathrm{max} \{I_\mathrm{opt}(\alpha)\}$, can be 
greatly enhanced when considering a nc-NESB (see Fig.~\ref{fig_n01} (a)).
It is also interesting to note that $I_\mathrm{opt}$ and $\theta_\mathrm{opt}$ is insensitive to the value of $\Delta$
except for very weak interaction strength (I),
which is in sharp constrast to the case of c-NESB~\cite{Segal2014:JPCA}.
This indicates a rather robust global heat current optimization $I_{\mathrm{max}}$ for $\{\alpha,\Delta,\theta\}$
once the bath temperatures are given, which might find practical utility in molecular junction engineering.

As $\Delta$ decreases, the transition between regime I and regime II occurs earlier and sharper (Fig.~\ref{fig_n02} (b)).
This transition finally disappears and there are only regime III and regime
IV left for a system with zero energy gap, which can be explained by Eq.~(\ref{eq_n04}). At $\Delta = 0$,
the contribution of the first term vanishes and only the second term survives, which is most pronounced when the two
coupling operators conmmute with each other, i.e. $\theta= 0$.
It can be expected that more diverse heat current behaviors will occur if we do not
constrain the second bath operator lying in the $x-z$ plane of the Bloch sphere 
and allow the rotation of both coupling operators.

% \section{Thermal Rectification}
\begin{figure}[t!]
	\includegraphics[width=0.75\columnwidth]{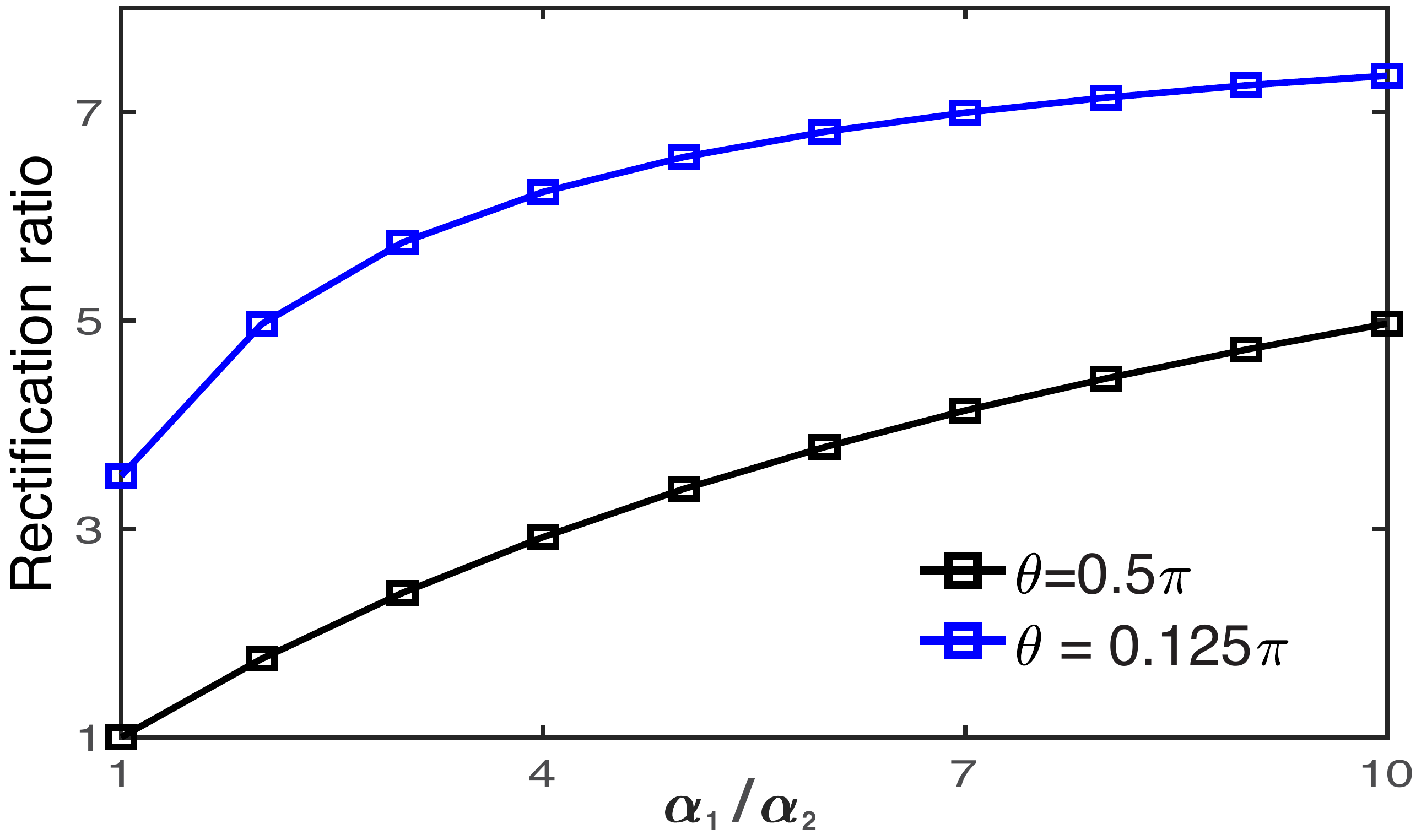}
	\caption{Thermal rectification for NESB with two different coupling operators: $\theta = 0.5\pi$ (black)
	and $\theta = 0.125\pi$ (blue), respectively. Rectification ratio is defined
	as the ratio between two values of heat current with the exchange of bath temperatures $T_{1}$ and $T_{2}$.
	We fix $\alpha_{1}=0.01$ and vary $\alpha_{2}$ to obtain different ratio of $\alpha_{1}/\alpha_{2}$.
	Other parameters are $T_{1} = 10$, $T_{2} = 1$ and $\Delta=0.5$.}
	\label{fig_n04}
\end{figure}

\textit{Thermal rectification.} 
Thermal rectification, which arises from the asymmetry in the total Hamlitonian,
offers rich possibilities to manipulate heat flow in nanosystems~\cite{Segal2005:PRL,BWLi:2012:RMP}.
In the c-NESB, the thermal rectification is usually realized by the asymmetry in coupling strength
~\cite{Segal2005:PRL,Segal2017:NJP}. Here we introduce an novel source of asymmetry,
non-commutative coupling operators between the system and two baths. Figure \ref{fig_n04} demonstrates the
thermal rectification ratio for the c-NESB ($\theta = 0.5\pi$) and a nc-NESB with $\theta = 0.125\pi$. 
A non-vanishing rectification occurs for the nc-NESB even at $\alpha_{1} = \alpha_{2}$, 
which is a pure quantum effect due to the asymmetry in coupling operators. 
More interestingly, the rectification ratio at $\theta=0.125\pi$
is significantly larger than that of the c-NESB for the entire parameter space.
This implies that two sources of asymmetry, coupling strength and coupling operators,
can work constructively to achieve optimal rectification.

% \section{Summary}
\textit{Summary.}
In this paper, we study heat transport properties of a generalized NESB with non-commutative system-bath coupling
operators and find unique transport properties different from those of the conventional NESB.
Scaling behaviors of the heat current with respect to the interaction strength
and the system energy gap are conspicuously altered when the two coupling operators do not commute, 
giving $I \sim \alpha^{2}$ in the weak coupling limit and
$I(\Delta \rightarrow 0^+)\neq 0$ in the adiabatic limit,
in sharp contrast to that $I \sim \alpha$ and $I(\Delta \rightarrow 0^+) \rightarrow 0$ for the conventional NESB.
These scaling relations can be explained analytically by the NE-PTRE.
Optimization for the heat current is performed using the extended HEOM, and four different regimes are distinguished. 
Given the termperature of two baths, a robust global optimal heat current can be obtained, 
independent to the system energy gap.
The heat current can be significantly enhanced with proper manipulation of the system-bath coupling operators. 
Asymmetry originated from the asymmetrical coupling strength and non-commutative coupling operators
can contribute constructively to thermal rectification, resulting in an enhanced rectification ratio.
The enhancement of heat current and thermal rectification due to non-commutative coupling
offer new and potentially advanced techniques for heat flow control.
We emphasize that these unusual transport properties reported in this paper 
are soley caused by the quantum effect of commutation and can also
be found in other nanoscale systems, including quantum heat engines and periodically driven systems
~\cite{Cao2019:arxiv,Alan2015:JPCL}.

% \begin{acknowledgments}
C. Duan, J. Liu and C.-Y Hsieh acknowledge the support from the Singapore-MIT Alliance for Research and Technology (SMART).
J. Cao is supported by NSF (grant no. CHE-1112825) and SMART. 
% \end{acknowledgments}


\begin{thebibliography}{300}
\bibitem{SM}
	See Suplememntary Materials for the details of the extended HEOM and the NE-PTRE.
	The $\alpha$ dependence on $I$ are also demonstrated with different $0 < \theta < 0.5\pi$.
\end{thebibliography}
\end{document}